\documentclass[prd,superscriptaddress,nofootinbib,amsmath,amssymb,aps,11pt]{revtex4-2}

\usepackage{bm}
\usepackage{amsfonts}
\usepackage{latexsym}
\usepackage[latin1]{inputenc}
\usepackage{graphicx}
\usepackage{amsmath}
\usepackage{palatino}
\usepackage{mathpazo}
\usepackage{booktabs}
\usepackage{dcolumn}

\def\jnl@style{\it}
\def\aaref@jnl#1{{\jnl@style#1}}

\def\aaref@jnl#1{{\jnl@style#1}}

\def\aj{\aaref@jnl{AJ}}                   % Astronomical Journal
\def\apj{\aaref@jnl{ApJ}}                 % Astrophysical Journal
\def\apjl{\aaref@jnl{ApJ}}                % Astrophysical Journal, Letters
\def\apjs{\aaref@jnl{ApJS}}               % Astrophysical Journal, Supplement
\def\apss{\aaref@jnl{Ap\&SS}}             % Astrophysics and Space Science
\def\aap{\aaref@jnl{A\&A}}                % Astronomy and Astrophysics
\def\aapr{\aaref@jnl{A\&A~Rev.}}          % Astronomy and Astrophysics Reviews
\def\aaps{\aaref@jnl{A\&AS}}              % Astronomy and Astrophysics, Supplement
\def\mnras{\aaref@jnl{Mon.~Not.~Roy.~Astron.~Soc.}}             % Monthly Notices of the RAS
\def\prd{\aaref@jnl{Phys.~Rev.~D}}        % Physical Review D
\def\prc{\aaref@jnl{Phys.~Rev.~C}}  % Physical Review C
\def\prl{\aaref@jnl{Phys.~Rev.~Lett.}}    % Physical Review Letters
\def\qjras{\aaref@jnl{QJRAS}}             % Quarterly Journal of the RAS
\def\skytel{\aaref@jnl{S\&T}}             % Sky and Telescope
\def\ssr{\aaref@jnl{Space~Sci.~Rev.}}     % Space Science Reviews
\def\zap{\aaref@jnl{ZAp}}                 % Zeitschrift fuer Astrophysik
\def\nat{\aaref@jnl{Nature}}              % Nature
\def\aplett{\aaref@jnl{Astrophys.~Lett.}} % Astrophysics Letters
\def\apspr{\aaref@jnl{Astrophys.~Space~Phys.~Res.}} % Astrophysics Space Physics Research
\def\physrep{\aaref@jnl{Phys.~Rep.}}      % Physics Reports
\def\physscr{\aaref@jnl{Phys.~Scr}}       % Physica Scripta
\def\commat{\aaref@jnl{Comm.~Math.~Phys.}}              % Communications in Mathematical Physics
\def\science{\aaref@jnl{Science}}               % Science
\def\cqg{\aaref@jnl{Classical Quant.~Grav.}}            % Classical and Quantum Gravity
\def\jpcs{\aaref@jnl{JPCS}}                                     % Journal of Physics Conference Series
\def\ijmpd{\aaref@jnl{Int.~J.~Mod.~Phys.~D}}                    % International Journal of Modern Physics D
\def\grg{\aaref@jnl{Gen.~Relat.~Gravit.}}               % General Relativity and Gravitation
\def\rpp{\aaref@jnl{Rep.~Prog.~Phys.}}          % Reports on Progress in Physics
\def\npa{\aaref@jnl{Nucl.~Phys.~A}}        % Nuclear Physics A
\def\lrr{\aaref@jnl{Living Rev.~Rel.}}                   % Living reviews in relativity
\def\jcap{\aaref@jnl{J.~Cosmology Astropart.~Phys.}}    % Journal of cosmology and astroparticle physics
\def\rmp{\aaref@jnl{Rev.~Mod.~Phys.}}   %Reviews of modern physics

\allowdisplaybreaks[1]
\begin{document}

	\title{Spontaneously scalarized black holes in dynamical Chern-Simons gravity: dynamics and equilibrium solutions}
	
	\author{Daniela D. Doneva}
	\email{daniela.doneva@uni-tuebingen.de}
	\affiliation{Theoretical Astrophysics, Eberhard Karls University of T\"ubingen, T\"ubingen 72076, Germany}

	\author{Stoytcho S. Yazadjiev}
	\email{yazad@phys.uni-sofia.bg}
	\affiliation{Theoretical Astrophysics, Eberhard Karls University of T\"ubingen, T\"ubingen 72076, Germany}
	\affiliation{Department of Theoretical Physics, Faculty of Physics, Sofia University, Sofia 1164, Bulgaria}
	\affiliation{Institute of Mathematics and Informatics, 	Bulgarian Academy of Sciences, 	Acad. G. Bonchev St. 8, Sofia 1113, Bulgaria}

	%%%%%%%%%%%%%%%%%%%%%%%%%%%%%%%%%%%%  DATE  %%%%%%%%%%%%%%%%%%%%%%%%%%%%%%%%%%%%
	%\date{\today}

	\begin{abstract}
   In the present paper, we construct spontaneously scalarized rotating black hole solutions in dynamical Chern-Simons (dCS) gravity by following the scalar field evolution in the decoupling limit. For the range of parameters where the Kerr black hole becomes unstable within dCS gravity the scalar field grows exponentially until it reaches an equilibrium configuration that is independent of the initial perturbation. Interestingly, the $\mathbb{Z}_2$ symmetry of the scalar field is broken and a strong maximum around only one of the rotational axes can be observed. The black hole scalar charge is calculated for two coupling functions suggesting that the main observations would remain qualitatively correct even if one considers coupling functions/coupling parameters producing large deviations from the Kerr solution beyond the decoupling limit approximation.
	\end{abstract}
	
	\maketitle
	
	\section{Introduction}
	An interesting class of modified gravity theories that reduce exactly to General relativity for spherically symmetric solutions and can deviate in the presence of a parity-odd source such as rotation is the dynamical Chern-Simons (dCS) gravity \cite{Jackiw:2003pm}. This modified theory possesses an additional dynamical (pseudo) scalar field coupled non-minimally to the Pontryagin topological invariant. Such a coupling arises naturally in loop quantum gravity \cite{Taveras:2008yf,Mercuri:2009zi,Ashtekar:1988sw} and  in effective field theories \cite{Weinberg:2008hq}.
	For a review on the subject, we refer the reader to \cite{Alexander:2009tp}. DCS gravity can be viewed as a special case of the extended scalar-tensor theories where the usual Einstein-Hilbert action is supplemented with all possible algebraic curvature invariants of second order with a dynamical scalar field nonminimally coupled to these invariants \cite{Berti:2015itd}. What makes the dCS gravity so interesting is the fact that the theory is parity-violating -- the deviations from general relativity occur only for systems that violate parity through the presence of a preferred axis.  Such systems are for example the isolated rotating black holes. The spinning black holes within  dCS gravity were studied perturbatively in a series of papers \cite{Yunes:2009hc,Konno:2009kg,Cambiaso:2010un,Yagi:2012ya,Stein:2014xba,Konno:2014qua,McNees:2015srl}. The only non-perturbative study of 
	spinning black holes in dCS gravity is that in \cite{Delsate:2018ome}  for a scalar field linearly coupled to the Pontryagin  invariant, while dCS black holes without $\mathbb{Z}_2$ isometry were considered in \cite{Cunha:2018uzc}.  The binary black hole merger of such black holes was examined in \cite{Okounkova:2019dfo}. 
	
	In the last years the  phenomenon of spontaneous scalarization of black holes attracted a lot of interest \cite{Stefanov:2007eq,Doneva:2010ke,Doneva:2017bvd,Silva:2017uqg,Antoniou:2017acq}. Most of the studies of this phenomenon were performed within the Gauss-Bonnet (GB) gravity and very little was done for the investigation of  the spontaneous scalarization in the case of 
	the dCS gravity. An exception is  \cite{Gao:2018acg} (see also \cite{Myung:2020etf}) where the authors have studied the tachyonic instability that triggers the spontaneous scalarization of the Kerr black hole within the quadratic dCS gravity. It was shown that the Kerr black hole becomes unstable under linear scalar perturbations in certain region of the parameter space. This is an indication that the Kerr black hole in the dCS gravity would get scalarized giving rise to a new non-Kerr black hole solution. In addition, as a toy model for rotation, the scalarization of Schwarzschild-NUT spacetime in dCS gravity was examined in \cite{Brihaye:2018bgc}. However, no spontaneously scalarized black hole solutions have been explicitly found/constructed in the dCS gravity. 
	
	The purpose of the present paper is to study the very dynamics of the spontaneous scalarization in dCS gravity, i.e. the process of forming scalarized black holes from Kerr black holes within dCS gravity  taking into account the non-linearities in the scalar field coupling similar to \cite{Doneva:2021dqn}. Since the scalarization dynamics of the  rotating black holes in its full generality and nonlinearity is an extremely difficult task,  in the present paper we consider the scalarization dynamics in the ``decoupling limit'' -- we numerically evolve the nonlinear scalar field equation on the fixed geometry background of a Kerr black hole, i.e.  we neglect the back reaction of the scalar field dynamics on the spacetime geometry. This approximation has proven to be very accurate in order cases of black hole scalarization, such as in GB gravity, if we keep the scalar charge small enough \cite{Doneva:2021dqn,Ripley:2019irj,Silva:2020omi,Witek:2018dmd}. 
	
	As end states of the dynamics, we also obtain the stationary scalarized black hole solutions in the dCS gravity in the  ``decoupling limit'' and study their properties. The solution of the full stationary field equations without approximations is very difficult since we have to deal with rotating solutions and field equations containing third-order derivative of the metric functions \cite{Delsate:2018ome}. That is why, despite the adopted approximation, the present studies can give us very valuable insight into the properties of scalarized black holes and allow us to compare with gravitational theories without parity violation, such as GB gravity.
	
	The paper is organized as follows. In Section II we present the basic background behind dCS gravity and derive the relevant scalar field evolution equation. Section III is devoted to the obtained numerical results starting with a description of the scalar field time evolution followed by an exploration of the properties of the equilibrium black holes. The paper ends with Conclusions.

	\section{Dynamical Chern-Simons gravity and scalar field evolution equation}	
	
	The action for the dCS gravity is given by 
	\begin{eqnarray}
		S= S=\frac{1}{16\pi}\int d^4x \sqrt{-g} 
		\Big[R - 2\nabla_\mu \varphi \nabla^\mu \varphi  
		+ 8\lambda^2 f(\varphi){\cal {}^{\star}R}{\cal R} \Big] ,\label{action}
	\end{eqnarray}
	where $R$ denotes the Ricci scalar with respect to the spacetime metric $g_{\mu\nu}$, $\nabla_{\mu}$ is the covariant derivative with respect to the spacetime metric $g_{\mu\nu}$ and  $f(\varphi)$ is the coupling function for the scalar field $\varphi$.  The  coupling constant $\lambda$ has  dimension of $length$ and ${\cal {}^{\star}R}{\cal R}$ denotes the Pontryagin invariant defined by ${\cal {}^{\star}R}{\cal R}={}^{\star}R_{\mu\nu\alpha\beta} R^{\mu\nu\alpha\beta} $, where  $R_{\mu\nu\alpha\beta}$ is  the Riemann tensor and ${}^{\star}R_{\mu\nu\alpha\beta}= \frac{1}{2}\epsilon_{\mu\nu\delta\gamma} R^{\delta\gamma}_{\,\,\,\,\,\alpha\beta}$ is its dual with $\epsilon_{\mu\nu\delta\gamma}$ being the 4-dimensional Levi-Civita tensor.
	
	The field equations derived from the action are 
	\begin{eqnarray}
	&&R_{\mu\nu} - \frac{1}{2}R g_{\mu\nu}  + 32\lambda^2 \left[ \nabla^{\alpha}f(\varphi) \epsilon_{\alpha\beta\gamma(\mu}\nabla^{\gamma} R_{\nu)}^{\;\;\beta} + \nabla^{\alpha}\nabla^{\beta}f(\varphi)  {}^{\star}R_{\beta(\mu\nu)\alpha} \right]
	\nonumber\\ &&= 2\nabla_\mu\varphi\nabla_\nu\varphi -  g_{\mu\nu} \nabla_\alpha\varphi \nabla^\alpha\varphi,  \\ \notag \\
	&&\nabla_{\alpha}\nabla^{\alpha}\varphi= - 2\lambda^2 \frac{df(\varphi)}{d\varphi} {\cal {}^{\star}R}{\cal R}.\label{SFE}
	\end{eqnarray}
	
	In the present paper  we are interested in  asymptotically flat spacetimes and  we shall consider  the case with $\varphi_{\infty}=0$. Without loss of generality we can impose on the coupling function $f(\varphi)$ the following conditions $f(0)=0$ and $\frac{d^2f}{d\varphi^2}(0)=\epsilon$ with $\epsilon=\pm 1$.
	In order for the spontaneous scalarization to occur we have to impose one more condition on the coupling function, namely $\frac{df}{d\varphi}(0)=0$.  
	When this condition is fulfilled  the  Kerr solution (with mass $M$ and angular momentum per unit mass $a$)
	\begin{eqnarray}\label{KerrM}
	ds^2=& -& \frac{\Delta -a^2\sin^2\theta}{\Sigma} dt^2 - 2a \sin^2\theta \frac{r^2 + a^2 - \Delta}{\Sigma} dt d\phi  
	+ \notag \\
	&+ &\frac{(r^2 + a^2)^2 - \Delta a^2 \sin^2\theta}{\Sigma} \sin^2\theta d\phi^2 + \frac{\Sigma}{\Delta} dr^2 + \Sigma d\theta^2,
	\end{eqnarray} 
	where $\Delta=r^2 - 2Mr + a^2$ and $\Sigma=r^2 + a^2 \cos^2\theta$,   is also a solutions to the  dCS field equations with a trivial scalar field $\varphi=0$. However,  for certain range of the parameters $M$, $a$ and $\lambda$   the Kerr solution  suffers from a tachyonic instability -- it becomes unstable within the dCS gravity as shown in \cite{Gao:2018acg}.  
	
	In the present paper using an approximate approach we show that  the exponential growth of the scalar field will last until the scalar field becomes large enough so that the nonlinear terms in the coupling function suppress the instability giving rise to a new stationary scalarized solution with a nontrivial scalar hair. 
	 As we have already commented the fully nonlinear dynamical problem is extremely difficult and that is why we shall base our study on an approximate approach which is free from heavy technical complications but preserves the leading role of the nonlinearity associated with the coupling function. In our approach we keep the spacetime geometry fixed and the whole dynamics is governed by the nonlinear equation for the scalar field. This dynamical approach is a very good approximation for example in the vicinity of the bifurcation point where the back reaction of the scalar field on the geometry is small, or away from the bifurcation but for coupling functions leading to relatively weak scalar field \cite{Doneva:2021dqn}. It was successfully applied in the case of binary black hole merger in GB gravity \cite{Silva:2020omi}. 
	 
	Following this approach we consider the nonlinear wave equation (\ref{SFE}) on the Kerr background geometry. In explicit form the equation (\ref{SFE}) takes the form  
	\begin{eqnarray} \label{eq:PertEq}
		&&-\left[(r^2 + a^2)^2 - \Delta a^2 \sin^2\theta\right] \partial^2_t \varphi + (r^2 + a^2)^2 \partial^2_x \varphi + 2r \Delta \partial_x \varphi - 4Ma r\partial_t\partial_{\phi_*}\varphi \nonumber \\ 
		&&+  2a(r^2 + a^2)\partial_x\partial_{\phi_*}\varphi  + \Delta\left[\frac{1}{\sin\theta} \partial_\theta(\sin\theta\partial_\theta\varphi) +   \frac{1}{\sin^2\theta}\partial^2_{\phi_*}\varphi \right] \\ 
		&& = -\lambda^2 \frac{192 a M^2\Delta}{\Sigma^5}  r\cos\theta  (3r^2- a^2\cos^2\theta)(r^2 -  3a^2 \cos^2\theta )\frac{df(\varphi)}{d\varphi}. \nonumber
	\end{eqnarray} 
	where we have introduced the tortoise coordinate $x$ and the new azimuthal coordinate $\phi_*$ defined by 
	\begin{eqnarray}
		dx= \frac{r^2+a^2}{\Delta} dr, \, \, \,   d\phi_*=d\phi + \frac{a}{\Delta} dr.
	\end{eqnarray} 
		
	We have also taken into account the explicit form of the  Pontryagin invariant for the metric \eqref{KerrM}, namely 
 	\begin{eqnarray} \label{eq:Potryagin}
 		{\cal {}^{\star}R}{\cal R}= \frac{96 a M^2}{\Sigma^6}  r\cos\theta  (3r^2- a^2\cos^2\theta)(r^2 -  3a^2 \cos^2\theta ).
 	\end{eqnarray} 
 	
 	We will conclude this section with comments on the conditions for scalarization (see also \cite{Gao:2018acg,Myung:2020etf}). For this purpose it is useful to consider the linearized version of eq. \eqref{eq:PertEq}, namely
 	 \begin{eqnarray}
 	 \nabla_{\alpha}\nabla^{\alpha}\delta\varphi= - 2\epsilon \lambda^2  {\cal {}^{\star}R}{\cal R}\delta \varphi,
 	 \end{eqnarray}
 	 where $\delta \varphi$ is the scalar field perturbation and $\epsilon = \frac{d^2f}{d\varphi^2}(0)$. The right-hand side of the equation gives rise to an effective scalar field mass that can be written in the form 
 	 \begin{equation} \label{eq:mu_eff}
	 	 \mu^2_{\rm eff} = - 2\epsilon \lambda^2  {\cal {}^{\star}R}{\cal R}.
 	 \end{equation}
 	 If $\mu^2_{\rm eff}<0$ a tachyonic instability is present that leads to scalarization of the Kerr black hole. The sign of $\mu^2_{\rm eff}$ is controlled by two factors. The first one is the sign of $\epsilon$ and the second one is the sign of the Pontryagin invariant. Taking the explicit form of ${\cal {}^{\star}R}{\cal R}$ in axial symmetry \eqref{eq:Potryagin} it is easy to show that eq. \eqref{eq:PertEq} is invariant under the change of the sign of $\epsilon$ combined with the change of the angle $\theta$ to $\pi-\theta$. Thus, the solutions with different $\epsilon$ are just mirrored one with respect to the other that gives us the freedom to consider only the case of  $\epsilon=1$ without loss of generality.

		\section{Results}
		
		\subsection{Scalar field coupling and domain of existence}
		Even though the adopted method can be very powerful and accurate in determining the final scalarized rotating black hole equilibrium solutions, the decoupling limit does no allow to rigorously prove the existence of solutions. In GB gravity for example the solutions can quickly disappear after the bifurcation point because the condition for the regularity of the scalar field and the metric functions at the black hole horizon is violated, and this behavior is strongly dependent on the choice of coupling function and the scalar field potential \cite{Doneva:2018rou,Silva:2018qhn,Macedo:2019sem,Doneva:2019vuh,Cunha:2019dwb,Collodel:2019kkx}. This violation can not be reproduced in our case if we assume that the background spacetime geometry remains the Kerr one. Luckily, such violation of the regularity conditions is not observed for the non-perturbative dCS black holes with linear coupling \cite{Delsate:2018ome} (moreover extremal black holes exist in dCS gravity \cite{Chen:2018jed}). This gives us the confidence that the domain of existence of the scalarized solutions will probably span from the bifurcation point all the way to the extremal limit and we will adopt this assumption in the calculations below, i.e. we will examine the scalarization of both slowly rotating Kerr black holes and near-extremal solutions. 
		
		In our studies we will consider more conservative and ``secure'' choices of coupling functions based on the studies of scalarized black holes in GB theory. In GB theory the simplest possible case that leads to scalarization $f(\varphi)=\varphi^2$ causes instabilities that can be cured either by adding an additional quartic term in the function $f(\varphi)$  \cite{Minamitsuji:2018xde,Silva:2018qhn} or by considering a self-interaction scalar field potential \cite{Macedo:2019sem}. Still the domain of existence, the stability, etc. is strongly dependent on the weight of these stabilizing terms, i.e. on the values of the associated additional parameters. Much safer and easier to handle numerically in GB gravity are the cases when the coupling has an exponential form, where the solutions are often stable in the whole domain of existence \footnote{Not considering of course the region where the hyperbolic character of the field equations is lost \cite{Blazquez-Salcedo:2018jnn}.} and almost all of the cases of rotating black holes in GB theory were calculated for such couplings \cite{Cunha:2019dwb,Herdeiro:2020wei,Berti:2020kgk} (with the exception of \cite{Collodel:2019kkx}). That is why, similar to the studies in GB theory, we will adopt the following two choices of the coupling function \cite{Doneva:2018rou,Doneva:2021dqn}
	 \begin{eqnarray} 
		&&f_{\rm I} (\varphi)= \frac{1}{2\beta} \left(1 - e^{-\beta \varphi^2}\right) , \label{eq:coupling12} \\
		&&f_{\rm II} (\varphi)= \frac{1}{\beta^2}\left(1 - \frac{1}{\cosh(\beta \varphi)}\right) , \label{eq:coupling21}
		\end{eqnarray} 
		where $\beta$ is a parameter. For scalarized GB black holes the value of $\beta$ practically controls the degree of scalarization -- the increase of $\beta$ suppresses the scalar field and leads to solutions close to the Kerr one \cite{Doneva:2018rou}. This is also fulfilled in dCS gravity and according to our numerical experiments $\beta=12$ leads to relatively weak scalar field where the decoupling limit is a good approximation even away from the bifurcation point. That is why we will adopt this value in the following studies.

		The domain of existence of scalarized solutions in dCS gravity, that is practically independent on the particular form of the coupling function as long as the condition $\frac{df}{d\varphi}(0)=0$ is fulfilled, can be found in \cite{Gao:2018acg}. In our simulation we have chosen to use the black hole mass $M$ as a normalization parameter. Thus the family of solutions is described by two parameters (assuming that $\beta$ is fixed) -- the normalized black hole angular momentum $a/M$ and the constant associated with coupling function $\lambda/M$. Thus, for a fixed $\lambda/M$ there exists a threshold  $a/M$ (a bifurcation point) above which the Kerr solution loses stability and gives rise to a new class of scalarized solutions. This threshold $a/M$ decreases with the increase of $\lambda/M$, i.e. for larger coupling constant slower rotating black holes can scalarize. In the limit when $\lambda/M \rightarrow 0$ the bifurcation point tends to $a/M \rightarrow 0$, i.e. there is no lower limit on $a/M$ for the development of scalarization \cite{Gao:2018acg,Myung:2020etf}. In addition, the growth time of the scalar field, i.e. the characteristic time required for the scalar field to develop from a small perturbation, tends to infinity at the bifurcation point and quickly decreases as the angular momentum or the coupling parameter is increased. Below we will first discuss the dynamics of scalarization followed by an investigation of the scalarized equilibrium black hole properties. 
		
		\subsection{Dynamics of the scalarization}
	 	The time evolution of the scalar field equations is performed with the numerical code developed in \cite{Doneva:2021dqn} for the case of GB gravity with the necessary adjustments to handle the Pontryagin scalar.  The boundary conditions we impose are the standard ones -- the scalar field should have the form of an ingoing wave at the black hole horizon and an outgoing wave at infinity.
	 	
	 	\begin{figure}
			\includegraphics[width=0.32\textwidth]{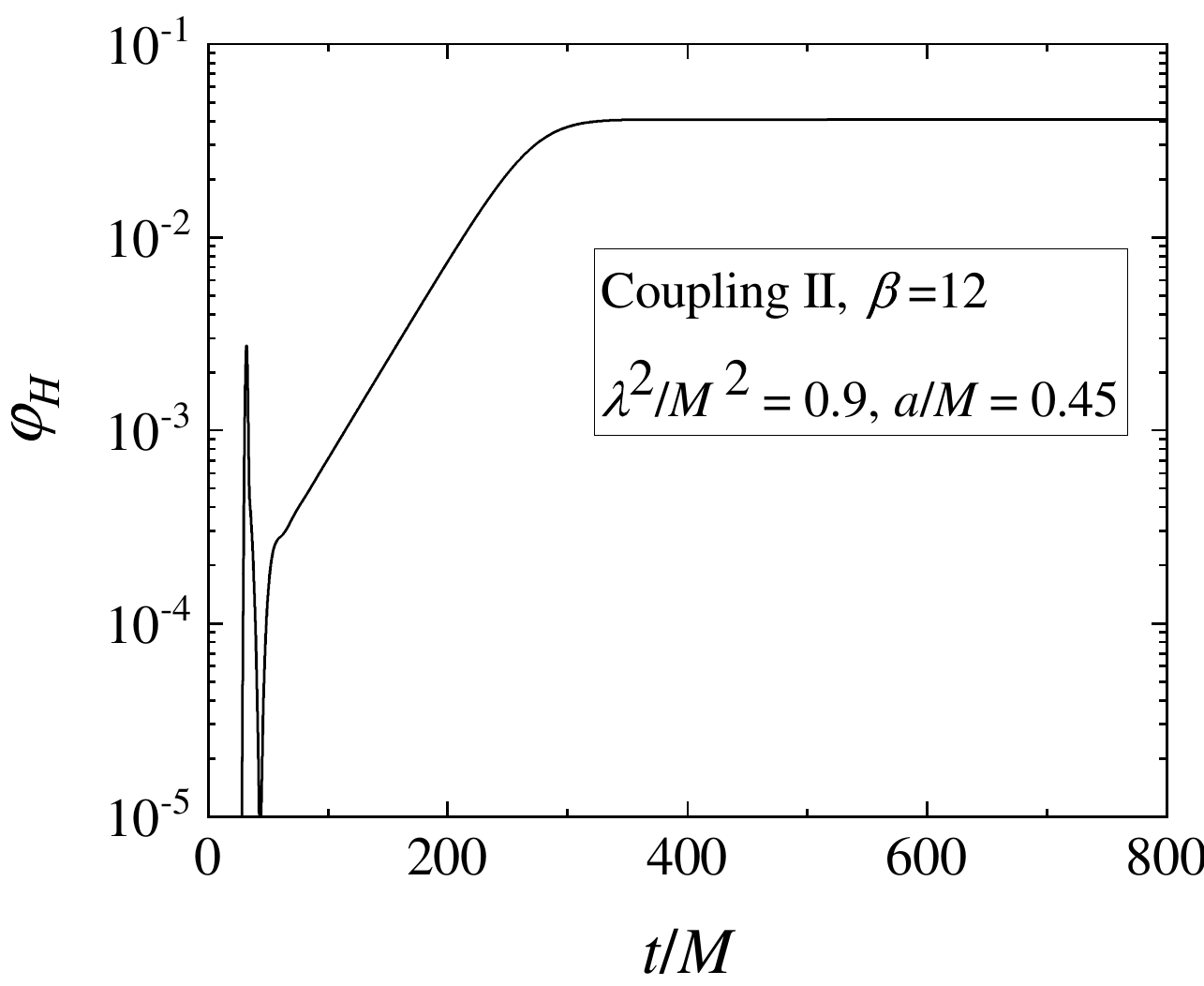}
			\includegraphics[width=0.32\textwidth]{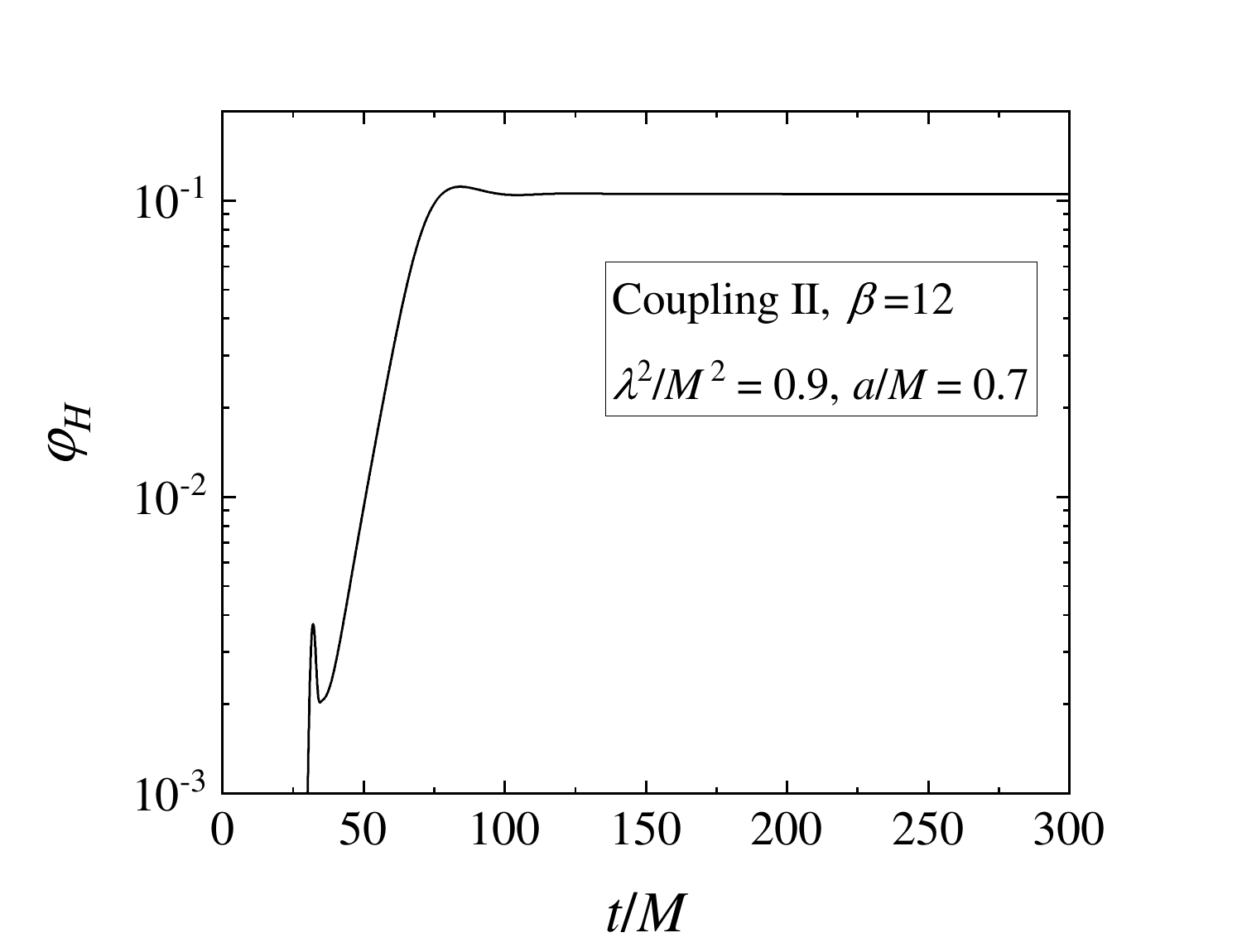}
			\includegraphics[width=0.32\textwidth]{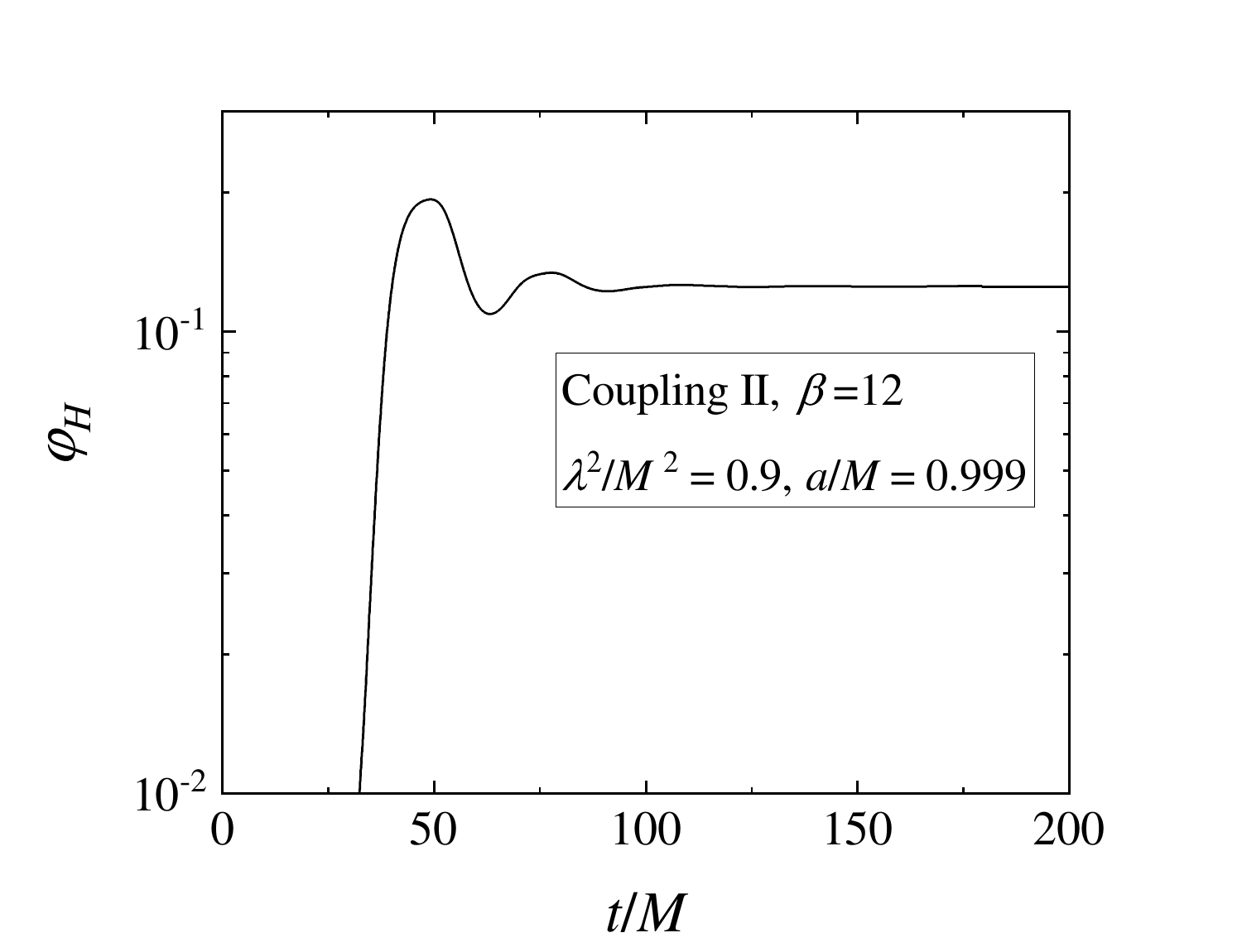}
			\caption{Time evolution of the scalar field on the horizon at $\theta=\pi/4$ for $\lambda^2/M^2=0.9$ and three different values of the angular momentum $a/M=0.45$, $a/M=0.7$ and $a/M=0.999$. The results are for the second coupling function \eqref{eq:coupling21} with $\beta=12$.}
			\label{fig:phiH_evolution}
		\end{figure}
		 \begin{figure}
		\includegraphics[width=0.45\textwidth]{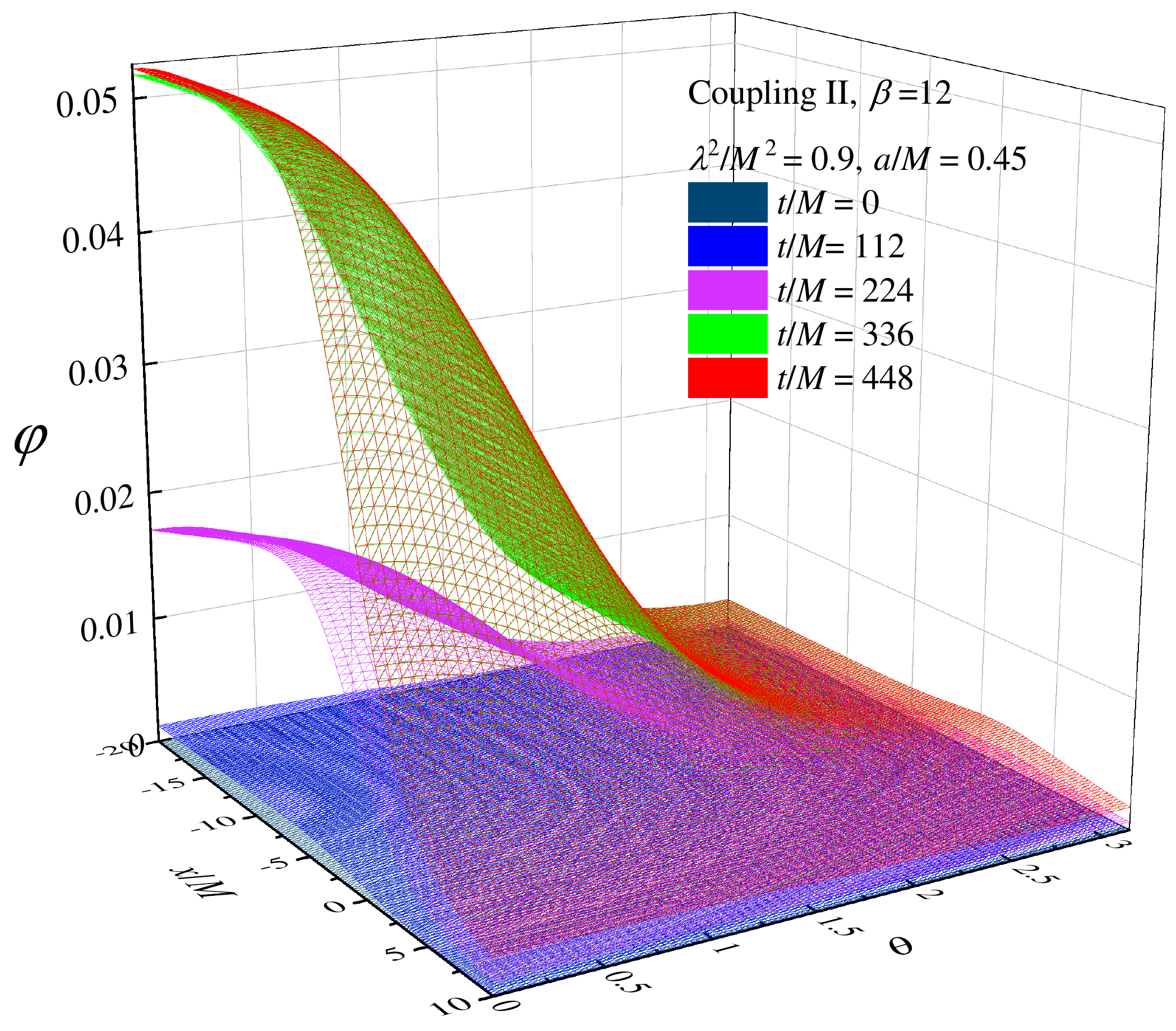}
		\includegraphics[width=0.45\textwidth]{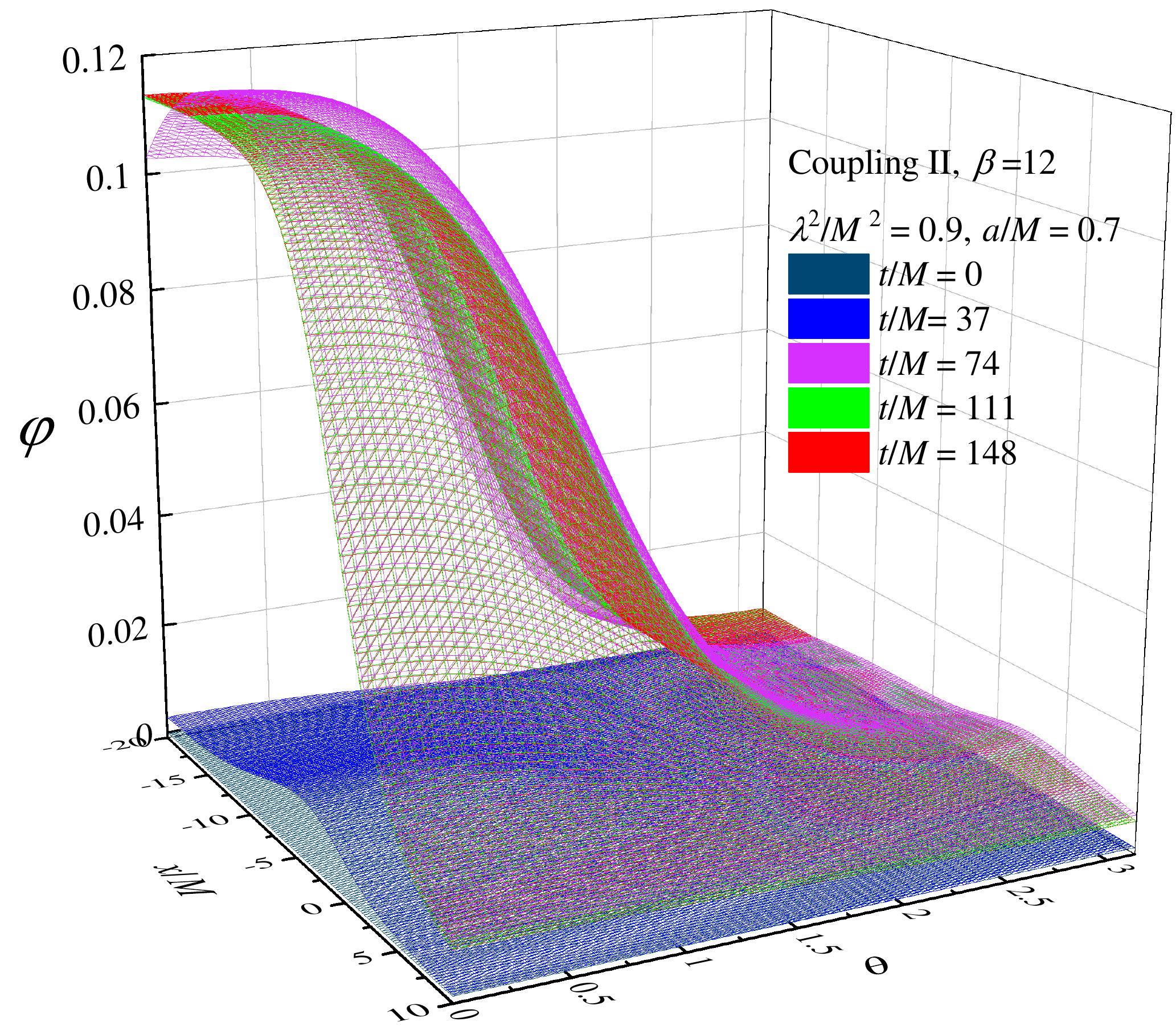}
		\includegraphics[width=0.45\textwidth]{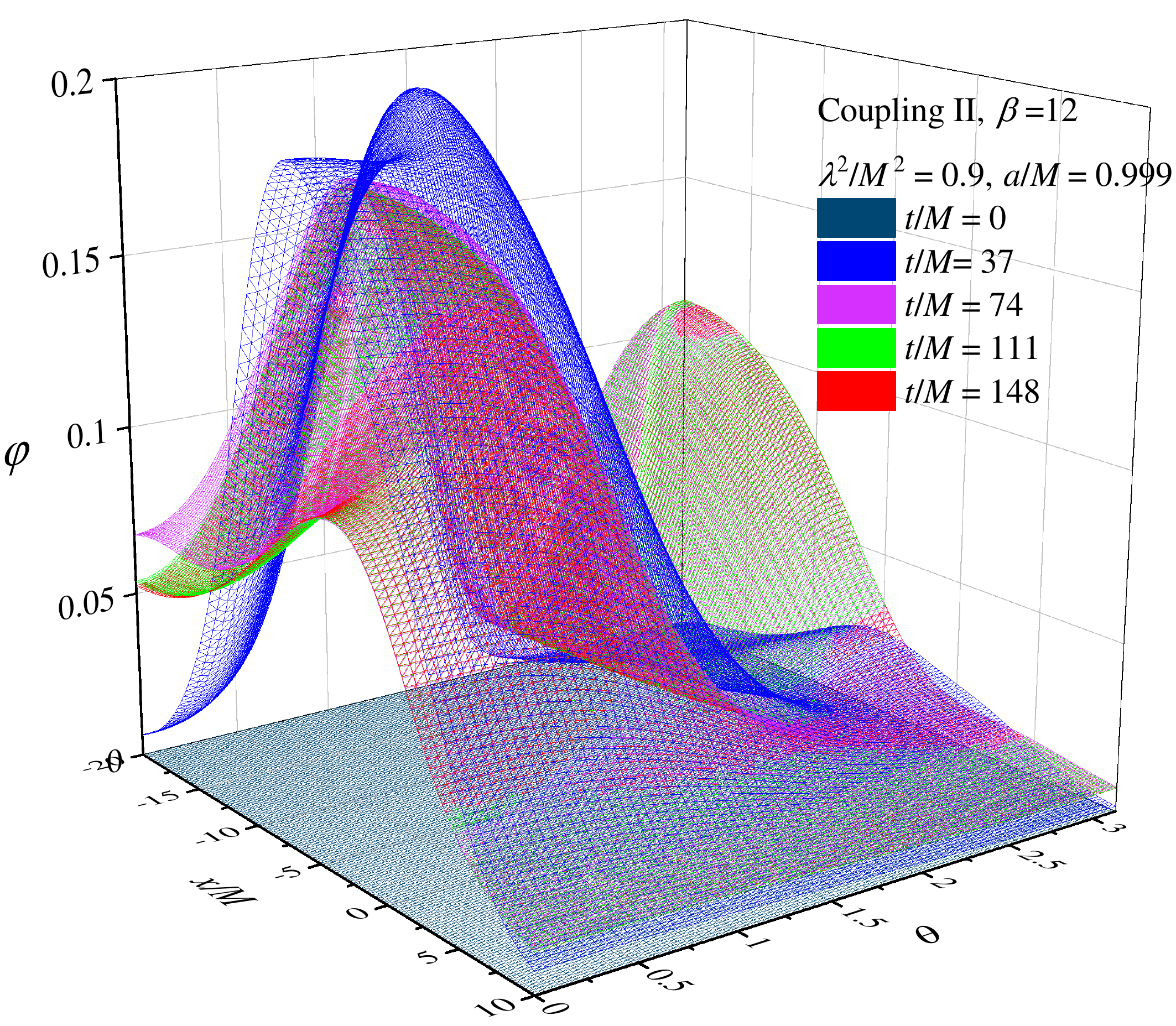}
		\caption{Snapshots of the two-dimensional scalar field profile at different times during the evolution for the same values of the parameters as in Fig. \ref{fig:phiH_evolution}. The last profile in time is always chosen to represent the final state after an equilibrium is reached.}
		\label{fig:psi_r_merged}
	\end{figure}

		The time evolution of the scalar field at the horizon is depicted in Fig. \ref{fig:phiH_evolution} for three different values of the normalized angular momentum, keeping fixed $\lambda^2/M^2=0.9$. For this $\lambda/M$ the bifurcation happens at $a/M=0.40$. The first model has $a/M=0.45$ (left panel) that is relatively close to the bifurcation point, the second has $a/M=0.7$ (middle panel) that is loosely speaking in the middle of the scalarization window, and $a/M=0.999$ (right panel) that is very close to the Kerr extremal limit. It is of course clear that if one takes into account the backreaction of the scalar field on the metric, this extremal limit will be different for Kerr and the scalarized black holes. Still, $a/M=0.999$ can give us good intuition of what can happen for near-extreme rotation. In all simulations, we have started with initial data in the form of a Gauss pulse located far outside the black hole horizon with an amplitude adjusted to be around two orders of magnitude smaller than the equilibrium scalar field that develops at late times. As the pulse reaches the near vicinity of the black hole horizon it starts to grow exponentially until it saturated to an equilibrium value, i.e. the black hole scalarizes. For models with low values of the angular momentum (left panel in Fig. \ref{fig:phiH_evolution}), a few oscillations are observed before the scalar field starts to grow exponentially. Close to the Kerr extremal limit (right panel in Fig. \ref{fig:phiH_evolution}) the growth time of the scalar field is so small that almost an immediate exponential increase is observed. Even more, clear oscillations with large amplitude are present around the equilibrium value of the scalar field before it settles to a constant.     
		
		Snapshots of the scalar field taken at different times during the evolution for the same black hole models are plotted in Fig. \ref{fig:psi_r_merged}. As observed also in Fig. \ref{fig:phiH_evolution}, for low angular momentum (for example $a/M=0.45$) more or less monotonic exponential increase of the scalar field is observed in the whole $(r,\theta)$ domain. Close to the extremal limit ($a/M=0.999$ in the figure) the evolution of the scalar field profile has quite different behavior -- first the scalar field increases until it reaches values larger than the equilibrium ones and afterwards it settles slowly to the equilibrium. Contrary to the low angular momentum case, the scalar field profile at intermediate times can be substantially different than the equilibrium ones. This can have interesting observational signatures for example in the emitted gravitational wave signal. In order to quantify the effect, though, one has to take into account also the backreaction of the scalar field on the spacetime geometry that is a very complicated task and we leave it for future studies.

	\subsection{Properties of the equilibrium black hole solutions}
	\begin{figure}
		\includegraphics[width=0.45\textwidth]{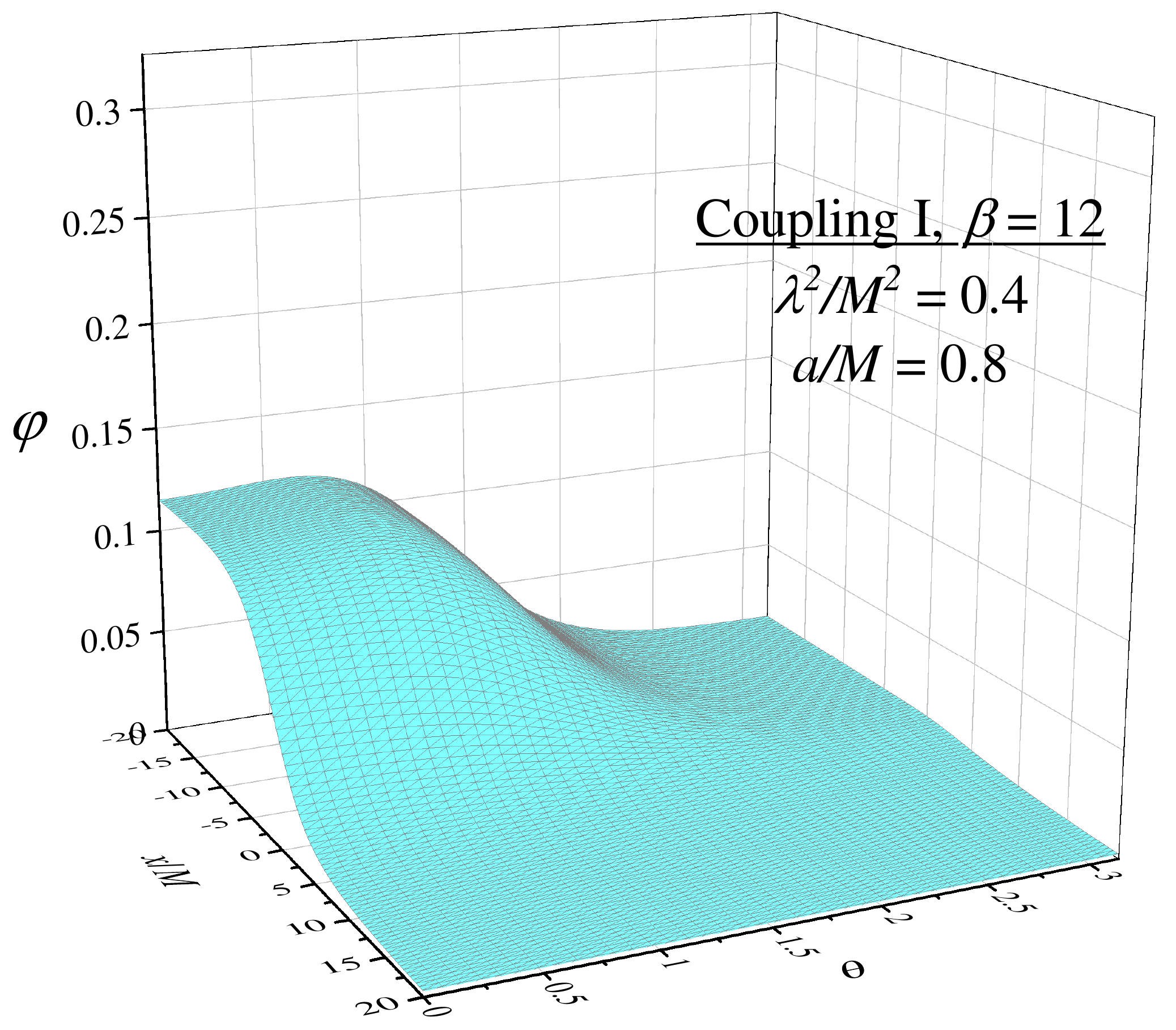}
		\includegraphics[width=0.45\textwidth]{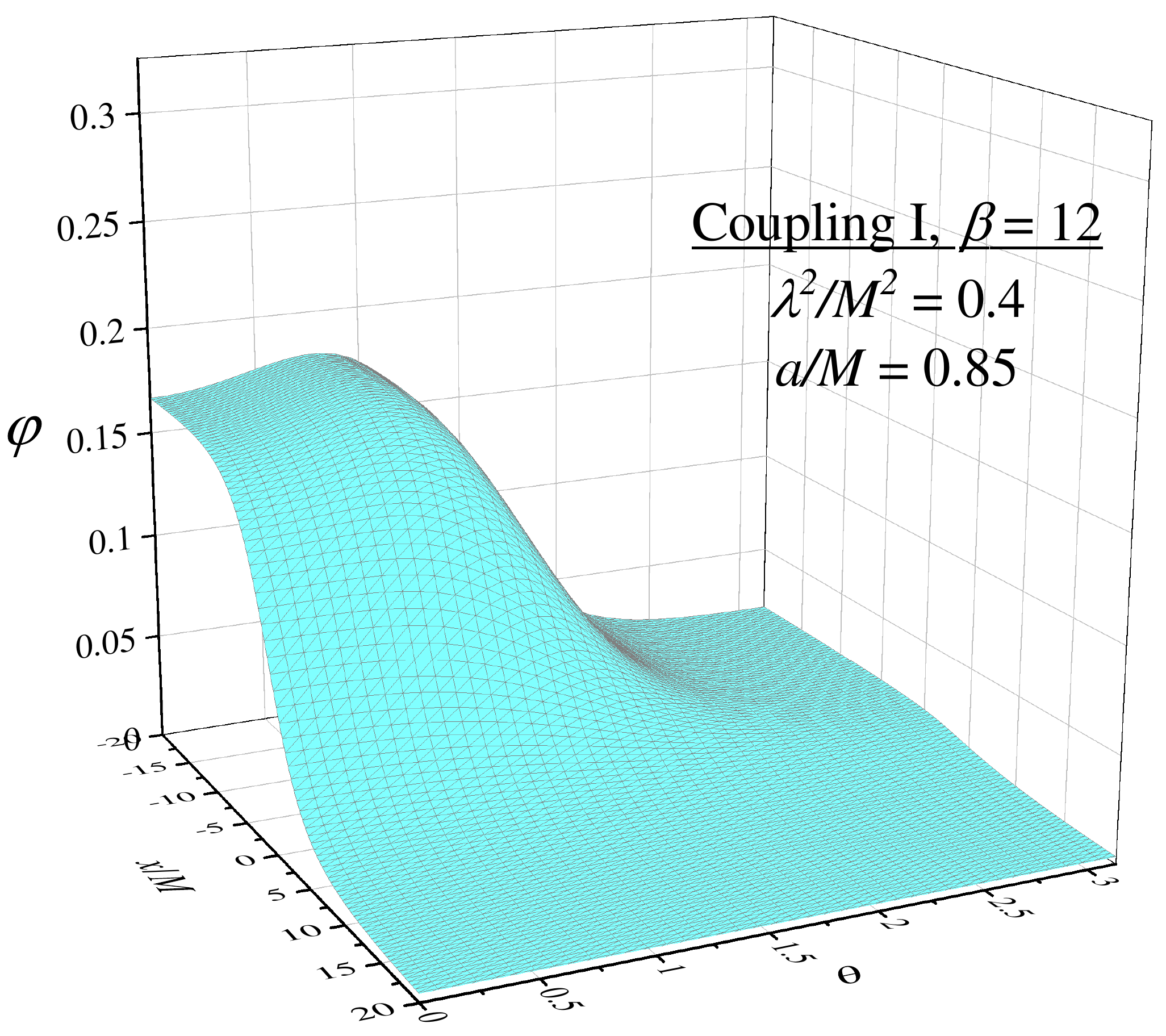}
		\includegraphics[width=0.45\textwidth]{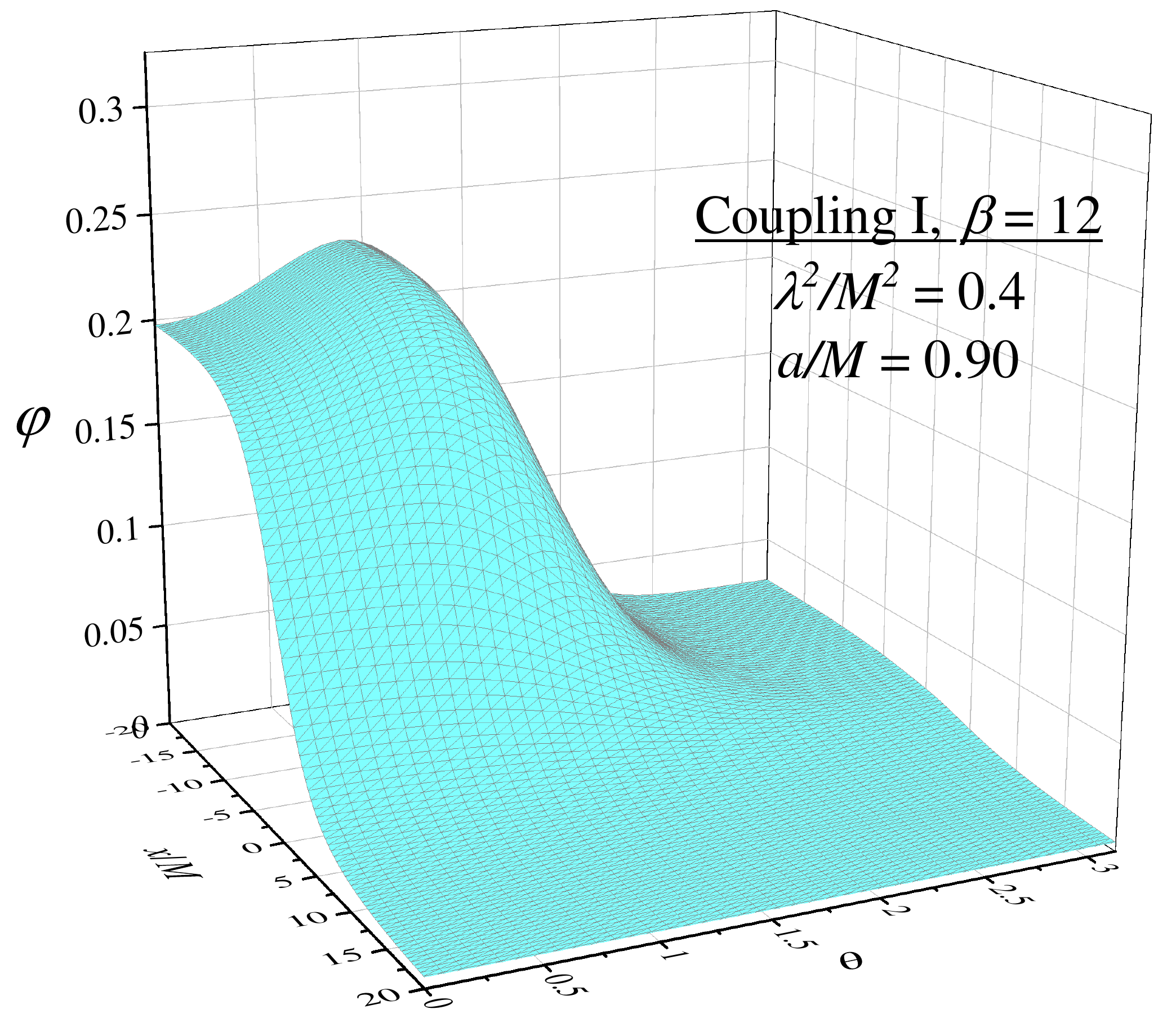}
		\includegraphics[width=0.45\textwidth]{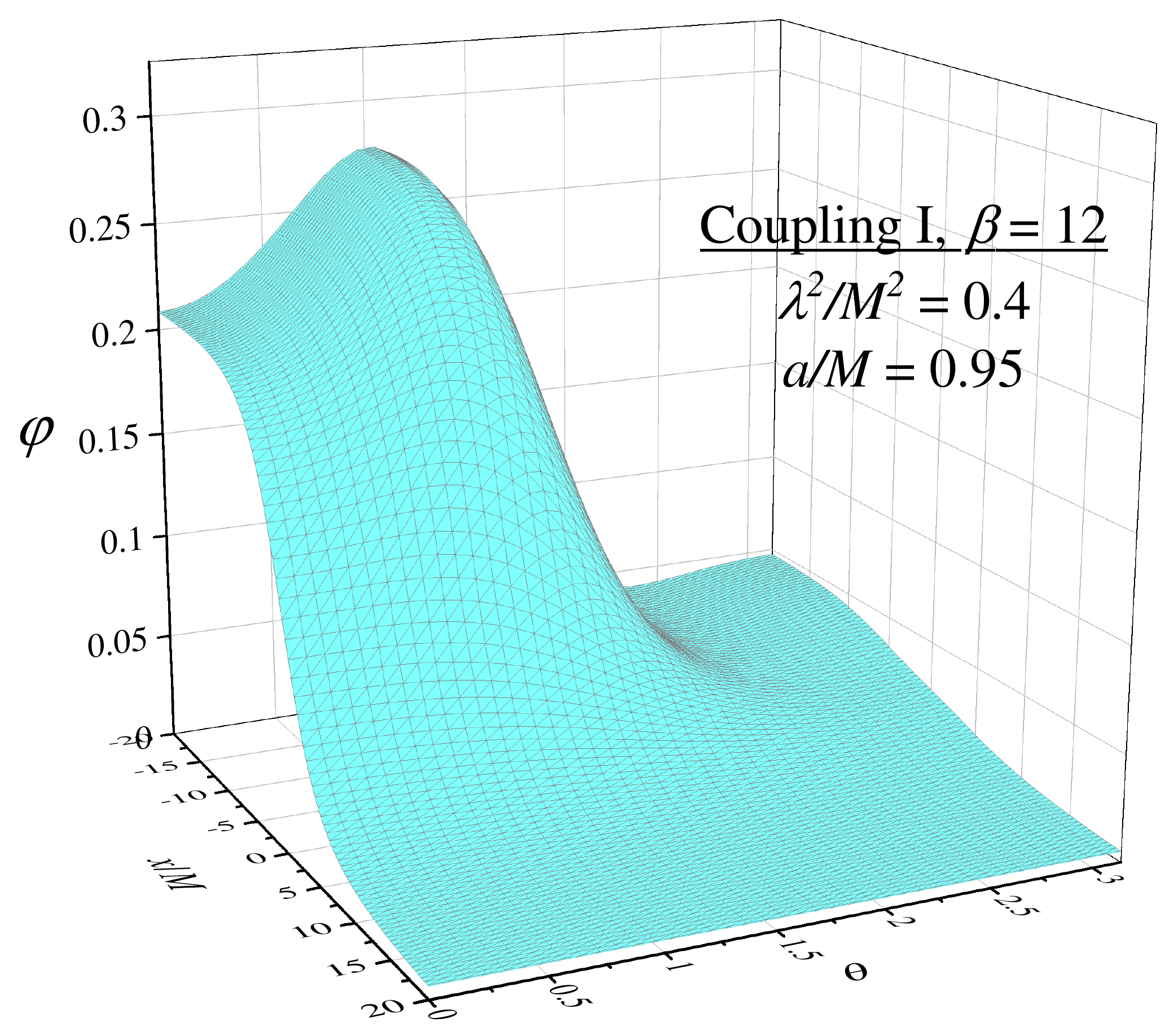}
		\includegraphics[width=0.45\textwidth]{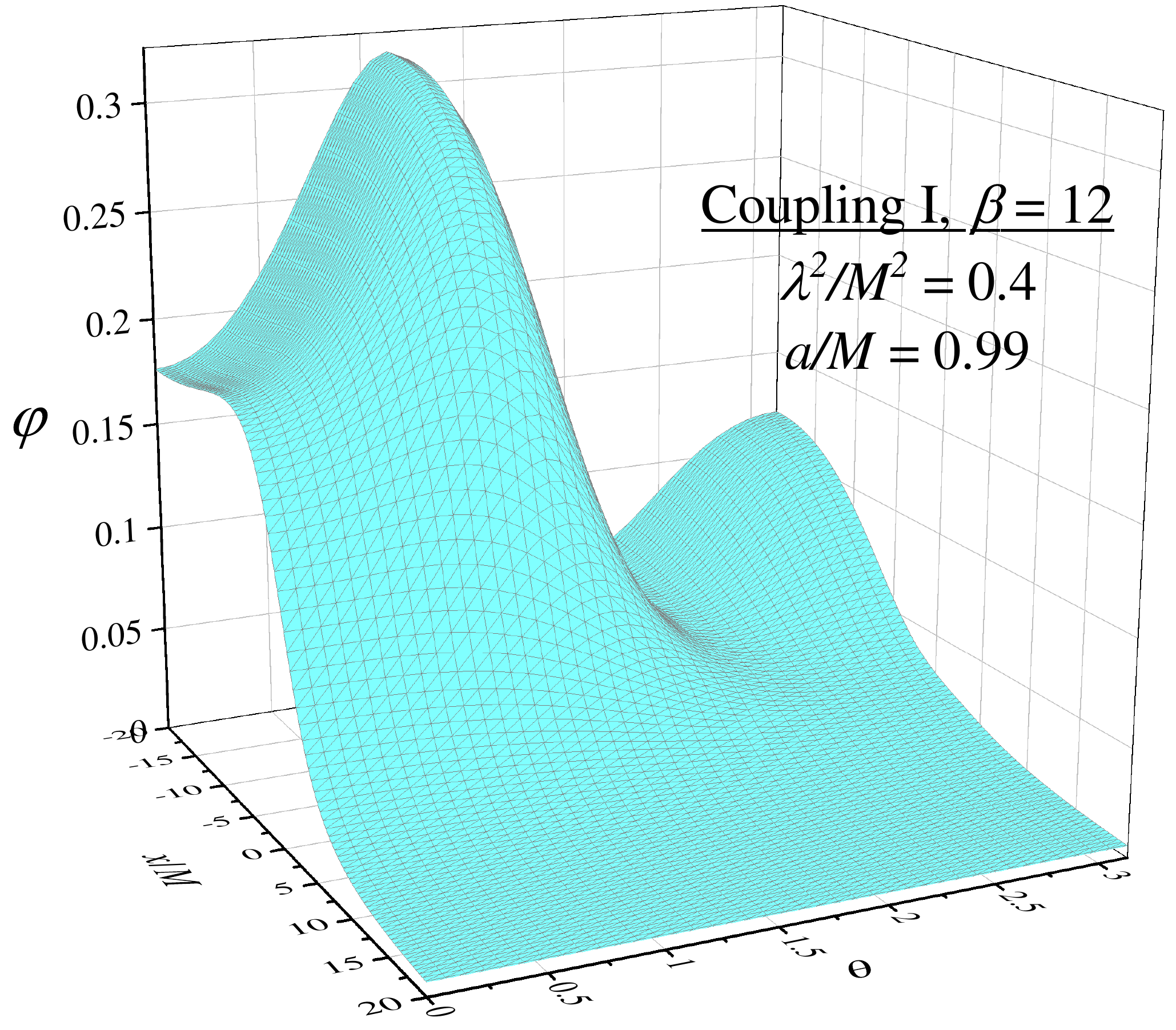}
		\caption{Some representative equilibrium scalar field profiles for fixed $\lambda^2/M^2=0.4$ and the first coupling \eqref{eq:coupling12} with $\beta=12$. Models with several different normalized angular momenta are plotted showing the change of the scalar field from models close to the bifurcation point all the way to the extremal limit.}
		\label{fig:psi_profiles}
	\end{figure}

	Even though the evolution discussed above can slightly change depending on the initial data, we have explicitly tested that the final equilibrium state is independent. A sample of solutions for a variety of parameters is plotted in Fig. \ref{fig:psi_profiles}. The most prominent feature one can notice is that they do not have a symmetry with respect to the plane $\theta=\pi/2$. This behavior is quite different compared for most of the example to dCS black holes with linear coupling \cite{Delsate:2018ome}, the GB black holes (with or without scalarization) \cite{Kleihaus:2015aje,Cunha:2019dwb}, or the rotating black holes in Einstein-Maxwell-dilaton gravity \cite{Herdeiro:2018wub} and is connected with the symmetries of the Chern-Simmons term appearing in the scalar field equation \eqref{eq:PertEq} if one considers coupling functions of the form \eqref{eq:coupling12} and \eqref{eq:coupling21}.
	
	The qualitative behavior of the scalar field for different $\theta$ can be understood if one examines more closely the criterion for scalarization $\mu^2_{\rm eff}<0$ where $\mu^2_{\rm eff}$ is defined by eq. \eqref{eq:mu_eff}. One should keep in mind that our discussion is specifically about the case of $\epsilon>0$. The Pontryagin invariant drops very rapidly as we go away from the black hole and that is why we will concentrate only on the region in the immediate vicinity to the black hole horizon. Clearly, the condition for the presence of tachyonic instability $\mu^2_{\rm eff}<0$ is $\theta$ dependent. The sign of the Potryagin invariant is controller by the $\cos\theta$ term in eq. \eqref{eq:Potryagin} for low and moderate values of the angular momentum. In this case $\mu^2_{\rm eff}<0$ only for $0\le\theta<\pi/2$. Clearly, once the condition for scalarization is fulfilled for some $\theta$ a nontrivial scalar field will develop for the whole black hole, but as the numerical simulations demonstrate, it is much stronger close to the $\theta=0$ rotational axis. For large $a/M$ (more precisely for $a/M>0.866$ if we consider the immediate vicinity of the black hole horizon) the $(3r^2- a^2\cos^2\theta)$ term  in eq. \eqref{eq:Potryagin} can change sign as well allowing for stronger scalarization close to the $\theta=\pi$ rotational axis. Indeed, as we see in the figure, a second maximum of the scalar field forms at  $\theta=\pi$ for very large rotational rates.

	\begin{figure}
		\includegraphics[width=0.45\textwidth]{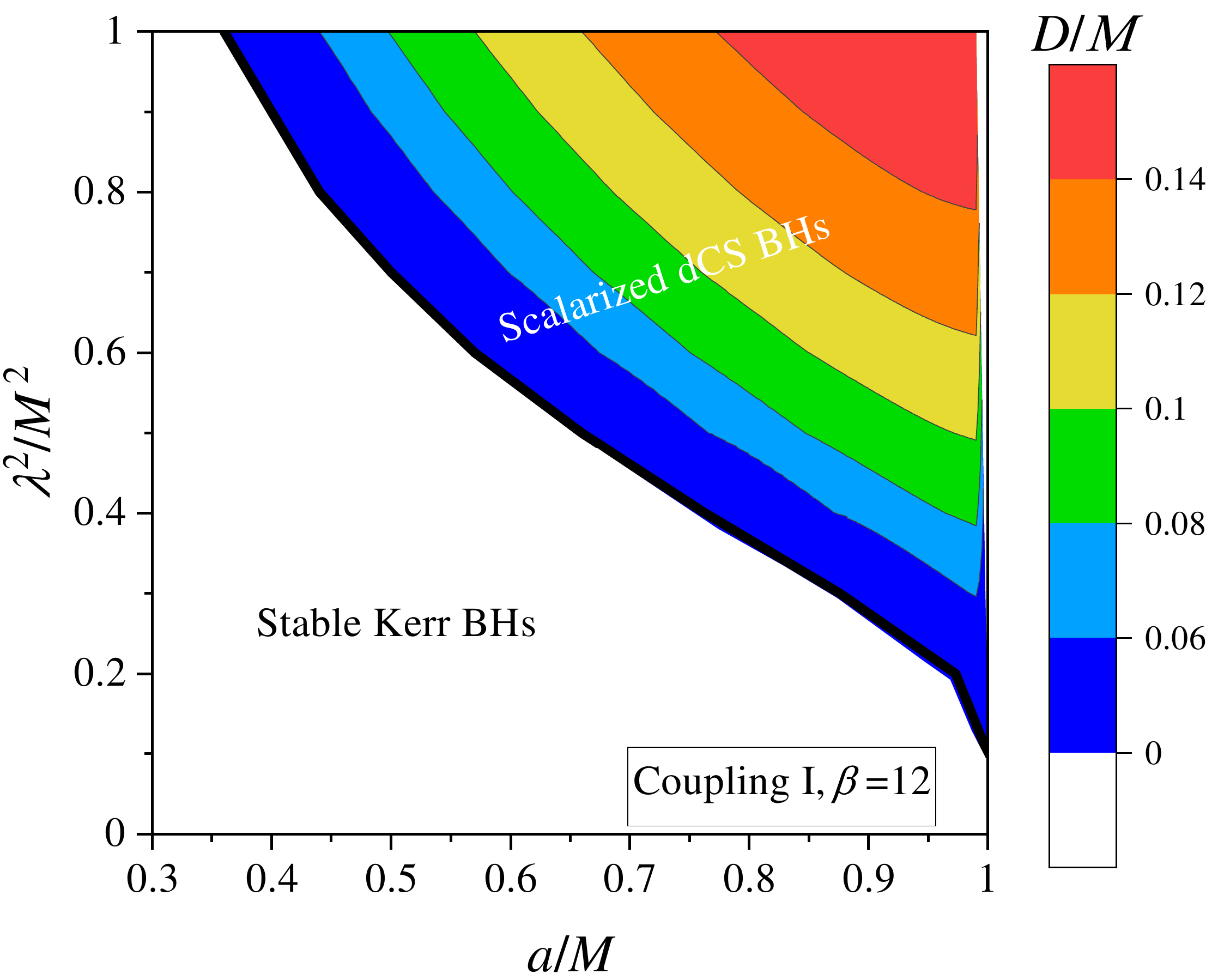}
		\includegraphics[width=0.45\textwidth]{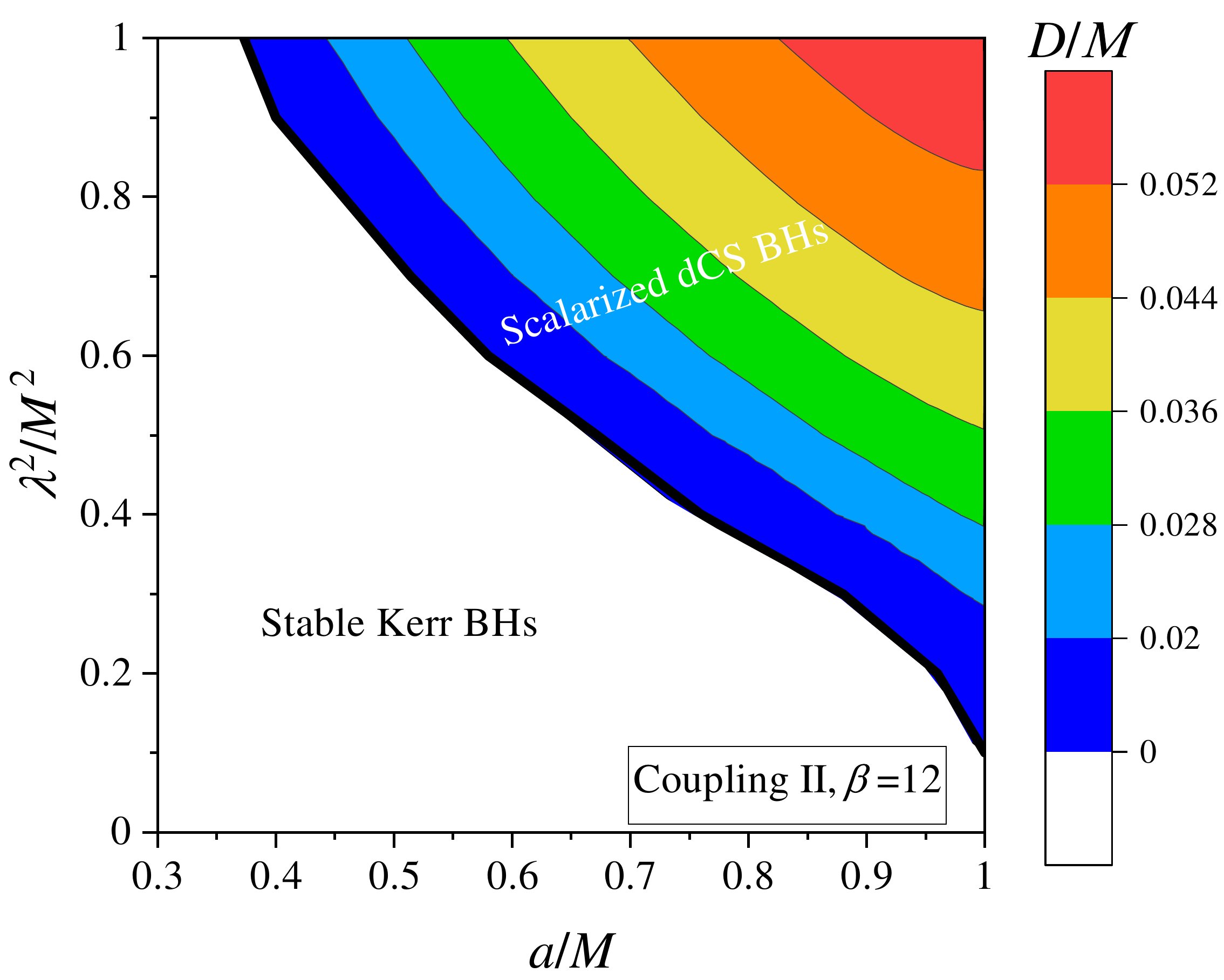}
		\caption{Contour plot of the normalized scalar charge $D/M$ as a function of the black hole angular momentum $a/M$ and coupling parameter $\lambda^2/M^2$. The results are for the both coupling functions \eqref{eq:coupling12} (left panel) and \eqref{eq:coupling21} (right panel) with $\beta=12$.}
		\label{fig:scalar_charge_lambda}
	\end{figure}
	
	Another quantity of interest is the scalar charge $D$. It is defined through the asymptotic of the scalar field at infinity. For the considered coupling functions, the leading order asymptotic has the form $\varphi \sim D/r$. This is very different from the case of Chern-Symons gravity with linear coupling where the scalar charge is zero \cite{Delsate:2018ome}. A contour plot of $D$ as a function of the angular momentum $a/M$ and the parameter $\lambda/M$ is given in Fig. \ref{fig:scalar_charge_lambda}. We have depicted the cases for both coupling functions. Close to the bifurcation point the accuracy of the calculations is somewhat reduced and thus the contour plots are not given in greater detail. The reason is purely computational. Close to the bifurcation point the growth time of the modes strongly increases and thus a much longer evolution time is required to see the development of accurate scalar field asymptotic. In addition, determining the scalar charge needs very good precision of the solution at large distances and any undesired reflected signal from infinity will spoil the procedure. Even though we have imposed outgoing wave boundary conditions, due to numerical inaccuracies there is always such undesired reflected signal\footnote{A more sophisticated method to eliminate the ingoing wave from infinity was implemented in \cite{Gao:2018acg}.} that eventually spoils the asymptotic behavior of the solution. Our way of completely ``eliminating'' the reflected signal is the simplest and most straightforward one -- we push the outer boundary to very large values and put an upper limit on the computational time so that the small reflected signal does not have the time to reach the point of extraction of the scalar charge. Clearly, if the scalar field grows more slowly longer computational time and thus larger grid spanning to larger numerical infinity is required, that greatly slows down the calculations.
	
	Clearly with the increase of $a/M$ or $\lambda/M$ the scalar charge increases monotonically. We have limited the figure up to $\lambda/M=1$ since for larger $\lambda/M$ the scalar charge quickly increases reaching large values for which the backreaction of the scalar field on the spacetime geometry can not be safely neglected anymore. An important property we can observe in Fig. \ref{fig:scalar_charge_lambda} is that the behavior of the scalar charge is qualitatively the same for both coupling functions and only the range of $D/M$ changes. Moreover, our studies show that the magnitude of $D$ scales with $\beta$, i.e. for the same mass models the increase of $\beta$ leads to a decrease of $D$ and vise versa. This suggests that the observations we have made in the present paper are relatively generic, more or less independent of the particular form of the coupling (as long as it leads to scalarization of course), and will remain approximately valid even if we consider the backreaction of the scalar field on the black hole metric.

	\section{Conclusion}
	In the present paper, we have studied non-perturbatively the dynamics of black hole scalarization in dynamical Chern-Symons gravity neglecting the backreaction of the scalar field on the spacetime metric. This approach has proven to give good results as long as the scalar field is kept small enough. Thus, we have limited ourselves to coupling functions and regions of the parameter space that fulfill this criterion. Apart from the study of the process of scalar field development, we have paid special attention to the properties of the newly formed equilibrium scalarized dCS black holes since such solutions were not obtained in the literature until now.
	
	The results show that close to the bifurcation point, i.e. for lower angular momentum of the black hole, the initial exponential increase of the scalar field is more or less monotonic until it settles to an equilibrium state.  Close to the extremal limit, though, the growth time of the scalar field is very large and as a result oscillations with relatively large amplitude around the equilibrium are observed. Thus the evolution towards this equilibrium state is not a monotonic one. Such behavior will have an influence on the metric perturbations and thus the gravitational wave emission, producing an interesting observational signature. In order to study this effect thoroughly, though, one has to take into account also the backreaction of the scalar field on the spacetime metric.
	
	The resulting equilibrium solutions at late times are also studied in detail. The scalar field profile is substantially different compared to most of the other solutions in dCS gravity (with linear coupling) or GB theory (including the case of scalarization). First of all, because of the symmetries of the Chern-Simons term in the scalar field equation, the $\mathbb{Z}_2$ symmetry of the scalar field is broken. In addition, for low and moderate rotational rates, the condition for scalarization is fulfilled only in one of the black hole ``hemispheres'' leading to a strong maximum of the scalar field at the corresponding rotational axis, and minimum at the other. Close to the extremal limit this behavior changes because the condition for the presence of tachyonic instability becomes more complicated, and thus two maxima (with different magnitude) appear at the two rotational axes. This behavior is very different than most of the beyond-Kerr black holes and can have various astrophysical implications. For example, the breaking of the symmetry with respect to the equator will have a clear signature for thick accretion discs, in the quasinormal mode spectrum, etc.
	
	 We have studied as well the scalar charge for a large domain of the parameters and for two coupling functions. The results show that the scalar charge gets stronger as we go to larger rotational rates and larger coupling parameter $\lambda$. We have limited our study to moderate values of $\lambda$, more precisely we focused on $\lambda/M<1$ in order to have a relatively small scalar field for which the decoupling limit approximation is satisfied, but our studies suggest that the scalar charge can increase significantly for larger $\lambda/M$. The qualitative behavior of the results is the same for the coupling function we have considered suggesting that the main observations we have made in the paper would remain  correct even if one considers coupling functions that lead to a stronger scalar field beyond the decoupling limit approximation.

	\section*{Acknowledgements}
	We would like to thank C. Herdeiro and E. Radu for reading the manuscript and the useful suggestions. DD acknowledges financial support via an Emmy Noether Research Group funded by the German Research Foundation (DFG) under grant
	no. DO 1771/1-1.  SY would like to thank the University of T\"ubingen for the financial support.  
	The partial support by the Bulgarian NSF Grant KP-06-H28/7 and the  Networking support by the COST Actions  CA16104 and CA16214 are also gratefully acknowledged. 
	
	%%%%%%%%%%%%%%%%%%%%%%%%%%%%%%%%%%%%%%%%%%%%%%%%%%%%%%%%%%%%%%%%%%%%%%%%%%%%%%%
	
	\bibliographystyle{apsrev4-2}
	\bibliography{references}

\end{document}